# A cyclic cosmological model based on the f(ρ) modified theory of gravity

Yaoming Shi

Dept. of Chemistry
University of California at Berkeley,
Berkeley, CA 94720
USA

Dept. of Optical Science and Engineering
Fudan University
Shanghai, 200433
China

yaoming_shi@yahoo.com

# Abstract

We consider FLRW cosmological models for perfect fluid (with rho as the energy density) in the frame work of the f(rho) modified theory of gravity [V. N. Tunyak, Russ. Phys. J. 21, 1221 (1978); J. R. Ray, L. L. Smalley, Phys. Rev. D. 26, 2615 (1982) ]. This theory, with total Lagrangian R-f(rho), can be considered as a cousin of the F(R) theory of gravity with total Lagrangian F(R)-rho. We can pick proper function forms f(rho) to achieve, as the F(R) theory does, the following 4 specific goals, (1) producing a non-singular cosmological model (Ricci scalar and Ricci tensor curvature are bounded), ; (2) explaining the cosmic early inflation and late acceleration in a unified fashion; (3) passing the solar system tests; (4) unifying the dark matter with dark energy. In addition we also achieve goal number (5): unify the regular matter/energy with dark matter/energy in a seamless fashion. The mathematics is simplified because in the f(rho) theory the leading terms in Einstein's equations are linear in second order derivative of metric wrt coordinates but in the F(R) theory the leading terms are linear in fourth order derivative of metric wrt coordinates.

# I. Introduction

The rapid development of observational cosmology started from 1990s shows that the universe has undergone two phases of cosmic acceleration. The first one is called cosmic early inflation [1,2,3,4,5] that occurred prior to the radiation domination (see [6,7,8] for reviews). The second accelerating phase started after the matter domination. The unknown component (dark energy) gave rise to this late time cosmic acceleration [9,10,11,12,13,14,15,16,17,18,19,20,21,22,23,24,25,26,27,28](see [29,30,31,32,33,34] for reviews).

Various theories are developed in an effort to explain the cosmic early inflation and late accelerated expansion. The $F(R)$ modified theories of gravity (see [35,36,37] for reviews) have recently [38] become one of the leading popular candidates in (1) producing a non-singular cosmology model $\left(0 < R^2 < \infty\right)$, $\left(0 < R_{\mu\nu}R^{\mu\nu} < \infty\right)$; (2) unifying the cosmic early inflation and late accelerated expansion in a continuous fashion; (3) passing the solar tests; (4) unifying the dark matter with dark energy.

In this paper we consider cosmological models based on the (less well known) $f(\rho)$ modified theories of gravity for perfect fluid [39,40]. We show that, like the $F(R)$ theories, the $f(\rho)$ theory can also accomplish the same 4 goals.

In addition we show that with $f(\rho)$ theory we can accomplish goal number (5): unifying the regular matter/energy with dark matter/energy in a seamless fashion.

One added benefit is that the mathematics is simplified in $f(\rho)$ theory. This is because the leading terms in Einstein's equations are linear in $\left(\partial_\lambda \partial_\kappa g_{\mu\nu}\right)$ in the $f(\rho)$ theory but are linear in $\left(\partial_\lambda \partial_\kappa \partial_\alpha \partial_\beta g_{\mu\nu}\right)$ in the $f(R)$ theories.

In section II we briefly review the $f(\rho)$ theory. In section III we consider FLRW cosmology and discuss the resulting Friedmann equations in two flavors (one kind is in terms of $\left(\dot{a}(t), a(t), \rho(a), p(\rho), f(\rho)\right)$ and the other kind is in terms of $\left(\dot{\rho}(t), \rho(t), p(\rho), f(\rho)\right)$. In section IV we consider a cyclic cosmological model and go through the checklist to see if we can accomplish 5 goals mentioned in the abstract. In section V we compare $f(\rho)$ theory with $f(R)$ theories, Chaplygin gas, NED etc. Section VI is conclusion.

# II. The $f(\rho)$ theory of gravity

In $f(\rho)$ theory of gravity for isentropic perfect fluid [39, 40], the Einstein equations and the energy-momentum density tensor, derived from a Hilbert-Einstein like action, $S=(16\pi)^{-1}\int\left(R-f(\rho)\right)\sqrt{-g}\,d^4x$, are given by (with units $G=c=1$):

$$R_{\mu\nu}-\tfrac{1}{2}g_{\mu\nu}R=8\pi T_{\mu\nu} \tag{2.1}$$

$$T_{\mu\nu}=\left(\rho_{eff}(\rho)+p_{eff}(\rho)\right)u_\mu u_\nu+p_{eff}(\rho)\,g_{\mu\nu} \tag{2.2}$$

Where the effective energy density and the effective pressure are given by:

$$\rho_{eff}(\rho)=f(\rho) \tag{2.3a}$$

$$p_{eff}(\rho)=-f(\rho)+(\rho+p)\partial_\rho f(\rho) \tag{2.3b}$$

In (2.3b) $\rho\geq 0$, $p\geq 0$ are the energy density and the pressure of the perfect fluid. Throughout this paper, they are always nonnegative. The effective equation of state parameter, $w_{eff}(\rho)$, is then given by:

$$w_{eff}(\rho)=\frac{p_{eff}(\rho)}{\rho_{eff}(\rho)}=-1+(\rho+p)\partial_\rho \ln f(\rho) \tag{2.4}$$

When $f(\rho)\equiv\rho$ we have $\partial_\rho f(\rho)\equiv 1$, $\rho_{eff}(\rho)\equiv\rho$, $p_{eff}(\rho)\equiv p$, $w_{eff}(\rho)\equiv p/\rho$ and (2.2) reduces to the standard expression for perfect fluid. The terms $\rho_{eff}(\rho)$ and $p_{eff}(\rho)$ may be considered as the pressure and energy density for an effective perfect fluid.

We would like to emphasize that the gravitational Lagrangian, $R$, is the same as in the Einstein's general relativity. Only the material Lagrangian is changed from $\rho$ to $f(\rho)$. We would also like to emphasize that the function $f$ in $f(\rho)$ theory is an arbitrary function, just like function $F$ in $F(R)$ theory.

Since there is no detailed derivation of the energy-momentum tensor (2.2)-(2.3) in [39], and the formulas for energy-momentum tensor in [40] are different from (2.2)-(2.3), we follow the standard text book procedure and provide a detailed derivation of the energy-momentum tensor (2.2)-(2.3) in Appendix A.

One benefit of using the concept of effective perfect fluid is that all the exact solutions in the literature for perfect fluid with nonzero and independent pressure are solutions of (2.1) and (2.2). For example, if $\rho\left(x^{\mu}\right)$ and $p\left(x^{\mu}\right)$ are not related by an equation of state and if $\left\{\rho\left(x^{\mu}\right), p\left(x^{\mu}\right), g_{\mu\nu}\left(x^{\lambda}\right)\right\}$ solves the traditional Einstein's equation for perfect fluid, then $\left\{\rho_{eff}\left(x^{\mu}\right), p_{eff}\left(x^{\mu}\right), g_{\mu\nu}\left(x^{\lambda}\right)\right\}$ is a solution as well. For given $f(\rho)$, we can then obtain $\rho = f^{-1}\left(\rho_{eff}\left(x^{\mu}\right)\right) \equiv \bar{\rho}\left(x^{\mu}\right)$,
$p = -\bar{\rho}\left(x^{\mu}\right) + \left(p_{eff}\left(x^{\mu}\right) + \rho_{eff}\left(x^{\mu}\right)\right)\left(\partial_{\bar{\rho}} f(\bar{\rho})\right)^{-1} \equiv \bar{p}\left(x^{\mu}\right)$.

We remark that the conservation of the enegy-momentum becomes:

$$\nabla_{\nu} T^{\mu\nu} = 0 \qquad \text{[because of Bianchi identity } \nabla_{\nu}\left(R^{\mu\nu} - \tfrac{1}{2} R g^{\mu\nu}\right) = 0 \text{ ]} \tag{2.5}$$

The conventional formula for the conservation of the enegy-momentum becomes the low-energy approximation of (2.5):

$$0 = \nabla_{\nu} T^{\mu\nu} \xrightarrow{\rho\to 0} 0 = \nabla_{\nu} T_0^{\mu\nu} \qquad T_0^{\mu\nu} = (\rho + p)\, u^{\mu} u^{\nu} + p\, g^{\mu\nu} \tag{2.6}$$

One consequence of using this effective perfect fluid concept is that the following energy conditions may or may not be satisfied in general.

The weak energy condition $\rho_{eff}(\rho) \geq 0, \rho_{eff}(\rho) + p_{eff}(\rho) \geq 0$; (2.7a)
The null energy condition $\rho_{eff}(\rho) + p_{eff}(\rho) \geq 0$; (2.7b)
The strong energy condition $\rho_{eff}(\rho) + p_{eff}(\rho) \geq 0, \rho_{eff}(\rho) + 3p_{eff}(\rho) \geq 0$; (2.7c)
The dominant energy condition $\rho_{eff}(\rho) \geq \left|p_{eff}(\rho)\right|$. (2.7d)

We remark that energy-momentum conservation in (2.5) is a different concept than various energy conditions in (2.7). (2.5) is an equality but (2.7a-d) are inequalities. (2.5) involves covariant-derivative but (2.7a-d) do not.

The criteria for the selection of $f(\rho)$, in our opinion, is $f(\rho) \approx \rho$ when $\rho$ is small. This condition is necessary for passing the solar system tests. The cosmological constant $\Lambda$ could be included in $f(\rho)$.

In table 1 below, we present blackbody radiation inspired choices of function $f(\rho)$ that, as we show later, can be used to solve the intrinsic singularity problems in GR when $\rho \to \infty$ and to make the $f(\rho)$ theory non-singular.

---

Table 1. Comparison of the perfect fluid Lagrangian $f(\rho)$ in General Relativity and the density distribution formula $f(\nu)$ in black-body radiation (for simplicity, unimportant constants are removed in various black-body radiation formulas)

| Perfect fluid Lagrangian in General Relativity | $f(\rho)$ | density distribution formula in black-body radiation | $f(\nu)$ |
|---|---|---|---|
| Einstein | $f(\rho)=\rho$ | Rayleigh-Jeans | $f(\nu)d\nu = \nu^2 d\nu$ |
| This work | $f(\rho)=\rho e^{-b\rho}$ | Wien | $f(\nu)d\nu = \beta h\nu^3 e^{-\beta h\nu} d\nu$ |
| This work | $f(\rho)=\dfrac{b\rho^2}{e^{b\rho}-1}$ | Planck | $f(\nu)d\nu = \dfrac{\beta h\nu^3}{e^{\beta h\nu}-1} d\nu$ |

The form $f(\rho)=b\rho^2\left(e^{b\rho}-1\right)^{-1}$ $\left(0<b^{-1}\propto\rho_{Planck}\right)$ of Table 1 is inspired by the radiation energy density distribution formula, $f(\nu)d\nu=\beta h\nu^3\left(e^{\beta h\nu}-1\right)^{-1}d\nu$, that Planck [41] derived in 1901 to solve the ultraviolet $(\nu\to\infty)$ divergence problem of blackbody radiation. The form $f(\rho)=\rho e^{-b\rho}$ in Table 1 is similar to Wien's distribution formula, $f(\nu)d\nu=\beta h\nu^3 e^{-\beta h\nu}d\nu$ ($h\nu >> \beta^{-1}=kT$). In this sense, the form of $f(\rho)=\rho$ in Einstein's original general relativity theory, resembles the Rayleigh-Jeans distribution formula, $f(\nu)d\nu = \nu^2 d\nu$ ($h\nu << \beta^{-1}=kT$).

We are amazed, as we show later, that Planck's magic black-body radiation formula would show its charm again, after over 100 years, in shedding some light on solving the UV divergence problem (intrinsic singularity inside of black hole or at naked singularity, etc) in Einstein's general relativity theory of gravity as well.

Later on we find out that the exponential in $f(\rho)=\rho e^{-b\rho}$ or $f(\rho)=b\rho^2\left(e^{b\rho}-1\right)^{-1}$ are not convenient for subsequent mathematical manipulation and rational functions like $f(\rho)=\rho(1+b\rho)^{-m}$ ($m$ is a positive integer) can do similar job in making $f(\rho)$ nonsingular.

From (2.1)-(2.3) we obtain:

$$R^2 = (8\pi)^2 \left(4f(\rho) - 3(p+\rho)\partial_\rho f(\rho)\right)^2 = (8\pi)^2 \left(\rho_{eff} - 3p_{eff}\right)^2 \tag{2.5a}$$

$$R^{\mu\nu}R_{\mu\nu} = (8\pi)^2 \left\{3\left(f(\rho) - (p+\rho)\partial_\rho f(\rho)\right)^2 + f^2(\rho)\right\} = (8\pi)^2 \left(3p_{eff}^2 + \rho_{eff}^2\right) \tag{2.5b}$$

So for given $p(\rho)$ as long as we pick function $f(\rho)$ such that:

$$\lim_{\rho\to\infty} \left|\rho_{eff}(\rho)\right| = \lim_{\rho\to\infty} \left|f(\rho)\right| < \infty \tag{2.6a}$$

$$\lim_{\rho\to\infty} \left|p_{eff}(\rho)\right| = \lim_{\rho\to\infty} \left|(\rho+p)\partial_\rho f(\rho) - f(\rho)\right| < \infty \tag{2.6b}$$

We can deduce from (2.5) that

$$0 \le R^2 < \infty \qquad\qquad 0 \le R^{\mu\nu}R_{\mu\nu} < \infty \tag{2.7}$$

The black-body radiation inspired formulas $f(\rho) = \rho e^{-b\rho}$ or $f(\rho) = b\rho^2\left(e^{b\rho} - 1\right)^{-1}$ are clearly satisfying the conditions (2.6) and can be used to make Ricci scalar curvature and the Ricci curvature tensor nonsingular. If we assume $\lim_{\rho\to\infty} p(\rho) = C\rho^k \quad (C, k > 0)$, then we may select $f(\rho) = \rho\left(1 + b^m\rho^m\right)^{-n} \quad (b, m, n > 0, mn > k)$ to make $R^2$ and $R^{\mu\nu}R_{\mu\nu}$ nonsingular as well. Equation (2.6) serve as one of the sufficient conditions for making the $f(\rho)$ theory of gravity nonsingular.

# III. The Friedmann equations

In Friedmann-Lemaitre-Robertson-Walker (FLRW) cosmology, the comoving (infalling) spherical symmetric line element is given by: [42]

$$ds^2 = g_{\mu\nu}\, dx^\mu\, dx^\nu = -dt^2 + a^2(t)\left(1 - kr^2\right)^{-1} dr^2 + a^2(t)\, r^2 \left(d\theta^2 + \sin^2\theta\, d\theta^2\right) \tag{3.1}$$

Where $a(t)$ is the scale factor and $t$ is the cosmological time. The energy density $\rho(t)$ and pressure $p(t)$ are assumed to be functions of $t$ as well. The Einstein's equations (2.1) then reduce to the Friedmann equations **[39]**:

$$\frac{\dot{a}^2+k}{a^2}=\frac{8\pi}{3}\rho_{eff}(\rho)\equiv\frac{8\pi}{3}f(\rho) \qquad \left(G_t^t=8\pi T_t^t\right) \tag{3.2a}$$

$$\frac{\ddot{a}}{a}=-\frac{4\pi}{3}\left(\rho_{eff}(\rho)+3p_{eff}(\rho)\right)\equiv\frac{8\pi}{3}f(\rho)-4\pi(p+\rho)\,\partial_\rho f(\rho) \qquad \left(G_t^t-G_r^r=8\pi T_t^t-8\pi T_r^r\right) \tag{3.2b}$$

Where $\dot{a}(t)\equiv\frac{d}{dt}a(t)$. From the conservation of the enegy-momentum $\nabla_\nu T^{\mu\nu}=0$, one nonzero condition remains [39]:

$$0=\left(3\frac{\dot{a}}{a}+\frac{\dot{\rho}}{\rho+p}\right)\partial_\rho f(\rho) \tag{3.3a}$$

Assuming that $\partial_\rho f(\rho)=0$ only occurs at isolated points, (3.3a) then reduces to

$$0=3\frac{\dot{a}}{a}+\frac{\dot{\rho}}{\rho+p}\ . \tag{3.3b}$$

Equations (3.2b) and (3.3b) are not independent; one can pick one or the other.

Given $f(\rho)$ and $p(\rho)$, equations (3.2a) and (3.3b) can be used to solve the time evolution for the scale factor $a(t)$ and energy density $\rho(t)$. Equation (2.4) for $w_{eff}(\rho(t))$ will then tell us what effective equation of state looks like at that $\rho(t)$.

We copied Table 2 from [43] to demonstrate what effective equation of state $w_{eff}(\rho(t))$ may look like at various stages of $\rho(t)$ (the last entry is from [44]). In Figure 3 below, we will show that the effective equation of state $w_{eff}(\rho(t))$ for a new cyclic universe model does vary from less than -1 to larger than +1 in a continuous fashion.

---

Table 2. Effective equation of state $w_{eff}(\rho(t))$ at various values of $\rho(t)$.

| $w_{eff}(\rho(t))$ | Behave like material name: |
|---|---|
| $<-1$ | Phantom energy |
| $-1$ | Cosmological constant |
| $-2/3$ | Domain wall |
| $-1/3$ | Cosmic string |
| 0 | Matter |
| $1/3$ | Radiation |
| $>1/3$ | Ultralight |
| $>1$ | Ultrastiff perfect fluid |

---

Using (3.3b) we obtain $3H = 3\dot{a}/a = -\dot{\rho}/(\rho + p)$ ($H(t)$ is Hubble function) and $a(\rho) = a_0 \exp\left(-\int^{\rho} (\rho + p(\rho))^{-1} d\rho\right)$ where $a_0$ is a constant of integration. Substituting these results into (3.1a), we can rewrite the Friedmann equation (3.1a) as:

$$H^2 = \frac{\dot{\rho}^2}{9(p(\rho)+\rho)^2} = \left[\frac{8\pi}{3} f(\rho) - \frac{k}{a_0^2} \exp\left(2\int^{\rho} (\rho + p(\rho))^{-1} d\rho\right)\right] \quad (3.4)$$

For spatially flat ($k = 0$) model, (3.4) reduces to:

$$H^2 = \frac{\dot{\rho}^2}{9(p(\rho)+\rho)^2} = \frac{8\pi}{3} f(\rho) \quad (3.5)$$

This is probably as far as we can go without specifying $f(\rho)$ and $p(\rho)$. We would prefer to work with Friedman equations in the format of (3.4) and (3.5) instead of the Friedman equations in the format of (3.2a) and (3,3b). This is because, as we show in the next section, that we do not need to specify the Equation of State $p(\rho)(\geq 0)$ to obtain major features of cosmology models when $f(\rho)$ is given.

We could bundle (3.2a) and (3.3b) together and write the Friedmann equations in the following different formats as:

$$\left\{H^2 + \frac{k}{a^2} = \frac{8\pi}{3} f(\rho), \quad 3(\rho + p(\rho))H + \dot{\rho} = 0\right\} \qquad \text{[ref.39, 1978]} \qquad (3.6a,b)$$

We would consider the Friedmann equations in terms of $(\dot{a}(t), a(t), \rho(a), p(\rho), f(\rho))$ as in (3.6) or in terms of $(\dot{\rho}(t), \rho(t), p(\rho), f(\rho))$ as in (3.4,3.5) as one of the major results of the $f(\rho)$ theory of gravity.

This set of Friedmann equations are clearly different from but also closely related to various (modified/generalized) Friedmann equations in the literature [45,46,47,48,49,50,51].

For example [45] considered a dark fluid with Equation of State like,

$$\bar{p}(\bar{\rho}) = -\bar{\rho} - \bar{f}(\bar{\rho}) \quad . \tag{3.7}$$

In notation of the current work, the Einstein's equations are given by

$$(8\pi)^{-1} G_{\mu\nu} = (\bar{\rho} + \bar{p})\, u_\mu u_\nu + \bar{p}\; g_{\mu\nu} \tag{3.8}$$

The Friedmann equations in this case become,

$$\left\{ H^2 + \frac{k}{a^2} = \frac{8\pi}{3}\bar{\rho}, \qquad 3\bar{f}(\bar{\rho})H + \dot{\bar{\rho}} = 0 \right\} \tag{3.9a,b}$$

We can clearly see that (3.9) is similar to but different from (3.6).

(i) If we identify $\bar{\rho} = \rho_{eff}(\rho) = f(\rho)$ and identify $\bar{p} = p_{eff}(\rho) = -f(\rho) + (\rho + p)\partial_\rho f(\rho)$ , then (3.9a) is identical to (3.6a), but (3.9b) is different from (3.6b). The function $\bar{f}$ and $f$ are related via, $\bar{f}\left(f^{-1}(\rho)\right) = (\rho + p(\rho))\partial_\rho f(\rho)$ .

(ii) If we identify $\bar{\rho} = \rho$, then (3.9b) is identical to (3.6b), but (3.9a) is different from (3.6a).

Other examples are considered in [46] where the Friedman equations and Equation of State go like,

$$\left\{ H^2 + \frac{k}{a^2} = \frac{8\pi}{3}\sum_i \rho_i, \qquad \dot{\rho}_i + 3(\rho_i + p_i)H = 0, \quad (i = 1, 2, 3...,) \right\} \tag{3.10a,b}$$

$$p_i = p_i(\rho_i) \quad . \tag{3.11}$$

We can also show that (3.10) is similar to but different from (3.6).

(i) If we identify $\sum_i \rho_i = \rho_{eff}(\rho) = f(\rho)$ and identify $\sum_i p_i = p_{eff}(\rho) = -f(\rho) + (\rho + p)\partial_\rho f(\rho)$ , then (3.10a) is identical to (3.6a), but (3.10b) is different from (3.6b).

(ii) If we identify $\rho_{i=1} = \rho$, then (3.10b) with $(i = 1)$ is identical to (3.6b), but (3.10a) is different from (3.6a).

# IV. A cyclic cosmological model

We now consider a concrete and spatially flat ($k=0$) cosmological model. We will show that our model does not rely on the signature of the 3D space curvature $k$ to make universe close, open, or flat anymore. The function $f(\rho)$ is assumed to be a 7-parameter M-shaped:

$$f(\rho)=\rho\left(1-\frac{\delta^n}{\rho^n}\right)\left(\frac{\left(\rho^m-\beta^m\right)^2+\gamma^{2m}}{\beta^{2m}+\gamma^{2m}}\right)\left(1-\frac{\rho^l}{\alpha^l}\right) \quad (3.5)$$

$$(l,m,n>0,1\geq n) \qquad (\alpha>>\beta>>\delta>0) \qquad (\gamma>0)$$

Function $f(\rho)$ has the following properties: $f(\rho)$ has two positive roots $(\rho=\delta,\alpha)$ and two complex-conjugated roots $\left(\rho_\pm=\left(\beta^m\pm i\gamma^m\right)^{1/m}\right)$. $f(\delta<\rho<\alpha)>0$. The constant $\gamma$ is selected such that $f(\rho)$ has a local minimum near $\rho=\beta$. Thus $f(\rho)$ is M-shaped.

This form of $f(\rho)$ is the combination of 3 forms that often appeared in the literature; namely the RS brane world [52,53] induced form (a) $f(\rho)=\rho(1-\rho/\alpha)$, the Cardassian expansion [54] form (b) $f(\rho)=\rho\left(1+\beta^m/\rho^m\right)$ $(0<m<1)$, and cosmology constant form (c) $f(\rho)=\rho-\delta$. The form (a) is used for getting rid of intrinsic singularity and producing bouncing solution; form (b) and (c) for explaining cosmic late accelerated expansion. We would like to mention that form $f(\rho)=\rho(1-\rho/\alpha)$ was already used in the original paper of $f(\rho)$ theory of gravity **[39]** for getting rid of the intrinsic singularity at $a(t)=0$ and providing a bouncing solution in the Oppenheimer-Snyder dust collapsing model [55].

We substitute (3.5) into (3.4) and also rewrite the result in a form mimic 1D Newtonian dynamics:

$$\frac{1}{2}\dot{\rho}^2+W(\rho)=E=0 \quad (3.6a)$$

$$W(\rho)=-12\pi\left(\rho+p(\rho)\right)^2\rho\left(1-\frac{\lambda^n}{\rho^n}\right)\left(\frac{\left(\rho^m-\beta^m\right)^2+\gamma^{2m}}{\beta^{2m}+\gamma^{2m}}\right)\left(1-\frac{\rho^l}{\alpha^l}\right) \quad (3.6b)$$

Equation (3.6) represents a particle moving in a 1D potential well $W(\rho)$ in classic mechanics.

We remark that the Firedmann equations expressed in terms of $\left(\rho(t), \frac{d}{dt}\rho(t)\right)$ like those on (3.4) and (3.6) are better suited for further discussion and treatment in $f(\rho)$ theory than those traditional ones expressed in terms of $\left(a(t), \frac{d}{dt}a(t)\right)$. This is because we do not need to specify function $p(\rho)$ yet in (3.4) and (3.6).

As shown in the Figure 1 below that potential well $W(\rho)$ is W-shaped (inverse of M-shape because of the negative sign in front of it). Inherited from $f(\rho)$, $W(\rho)$ also has positive roots $(\rho = \delta, \alpha)$ and two complex-conjugated roots $\left(\rho_{\pm} = \left(\beta^m \pm i\gamma^m\right)^{1/m}\right)$. $W(\delta < \rho < \alpha) < 0$. This is the reason we would name it $W(\rho)$ instead of $V(\rho)$ or $U(\rho)$ in this paper.

Figure 1. History of cosmology for a cyclic universe model: From big band to cosmic early inflation to cosmic late accelerated expansion to big stop. Numerical constants are $l = m = n = 1$, $p = w\rho, w = 0$, $\alpha = 10, \beta = 6, \gamma = 1, \lambda = 10^{-3}$. The points A, B, C, D, and E are defined in the text below.

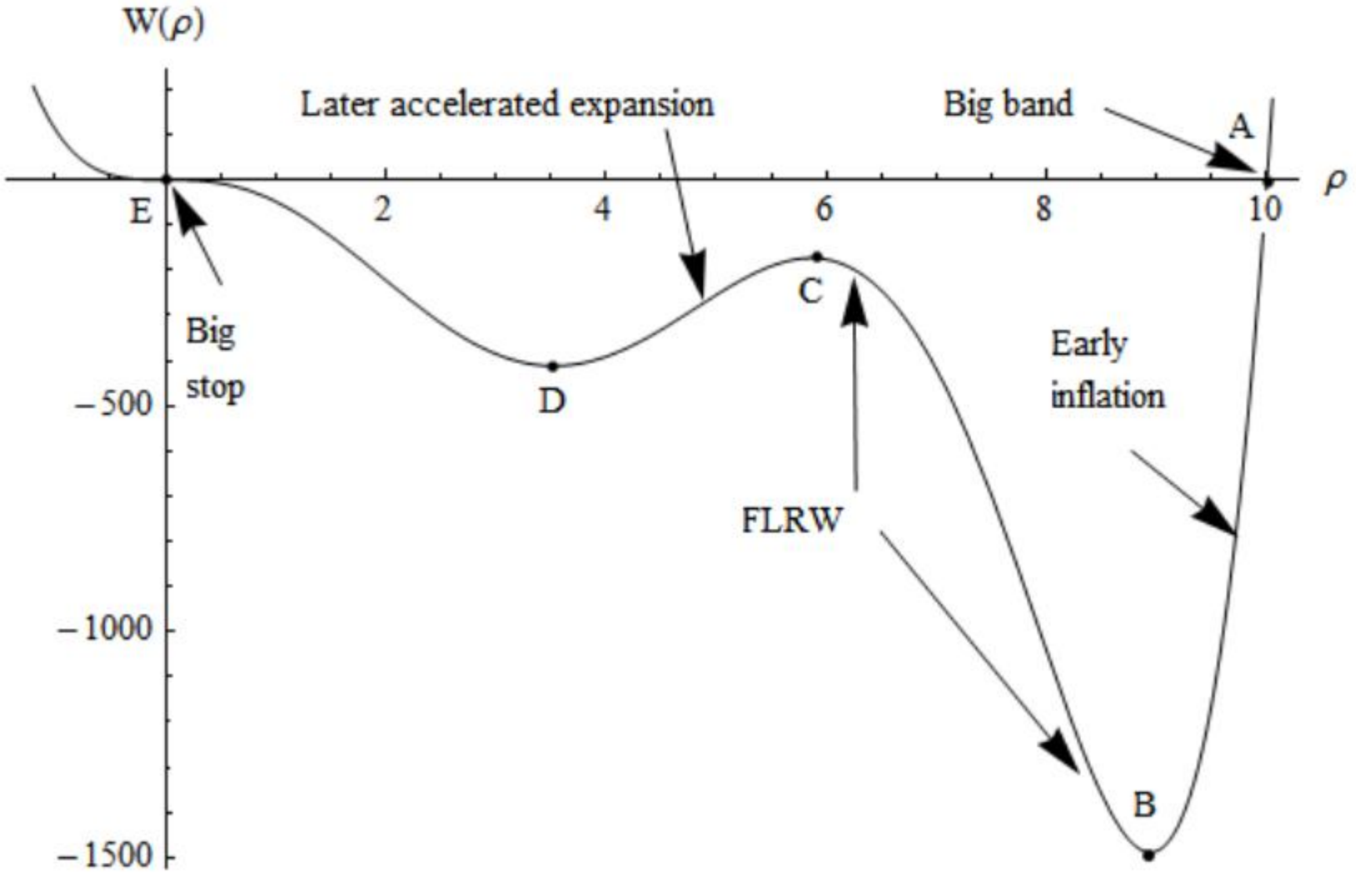

Without specifying $p(\rho)$ and solving the Fridmann equations (3.6), we can already illustrate the main features of this cosmological model. The big-bang starts at $\left(\rho(t=0)=\alpha>>1,\dot{\rho}(t=0)=0,W=0\right)$ (point A) as shown Figure 1. The top of the middle hump (point B) is near $(\rho=\beta)$, the early cosmic inflation happened roughly from $(\rho=\alpha)$ (point A) to $\left(\rho=\frac{1}{2}(\alpha+\beta)\right)$ (point B). The universe expansion described by the original FLWR model is roughly between $\left(\rho=\frac{1}{2}(\alpha+\beta)\right)$ (point B) to $(\rho=\beta)$ (point C). We are currently undergoing an accelerated expansion probably in the range from $(\rho=\beta)$ (point C) to $\left(\rho=\frac{1}{2}\beta\right)$ (point D). The universe will stop expansion at $\left(\rho(t=T)=\delta<<1,\dot{\rho}(t=T)=0,W=0\right)$ (point E) and turn itself around and start the big crunch and go all the way back to end the big crunch at $\left(\rho(t=2T)=\alpha,\dot{\rho}(t=2T)=0,W=0\right)$ (point A). It then starts a new big bang - big crunch process. So this cosmological model is a cyclic universe model. This model also covers the eras (like bouncing at both ends which are) beyond the early cosmic inflation and late cosmic accelerated expansion.

Figure 2. The plot of effective equation of state $w_{eff}(\rho)$ vs. $\rho$ for the M-shaped $f(\rho)$ of (3.7) Numerical constants are $l=m=n=1$, $p=w\rho, w=0$, $\alpha=10,\beta=6,\gamma=1,\lambda=10^{-3}$ . For convenience, we also indicate in this figure the points A, B, C, D, and E that are shown in Figure 1 and defined in text below. The two dashed lines are for $w_{eff}(\rho)=-1$, the phantom cross; and $w_{eff}(\rho)=+1$, the cross over to ultrastiff perfect fluid (see the last entry of Table 2).

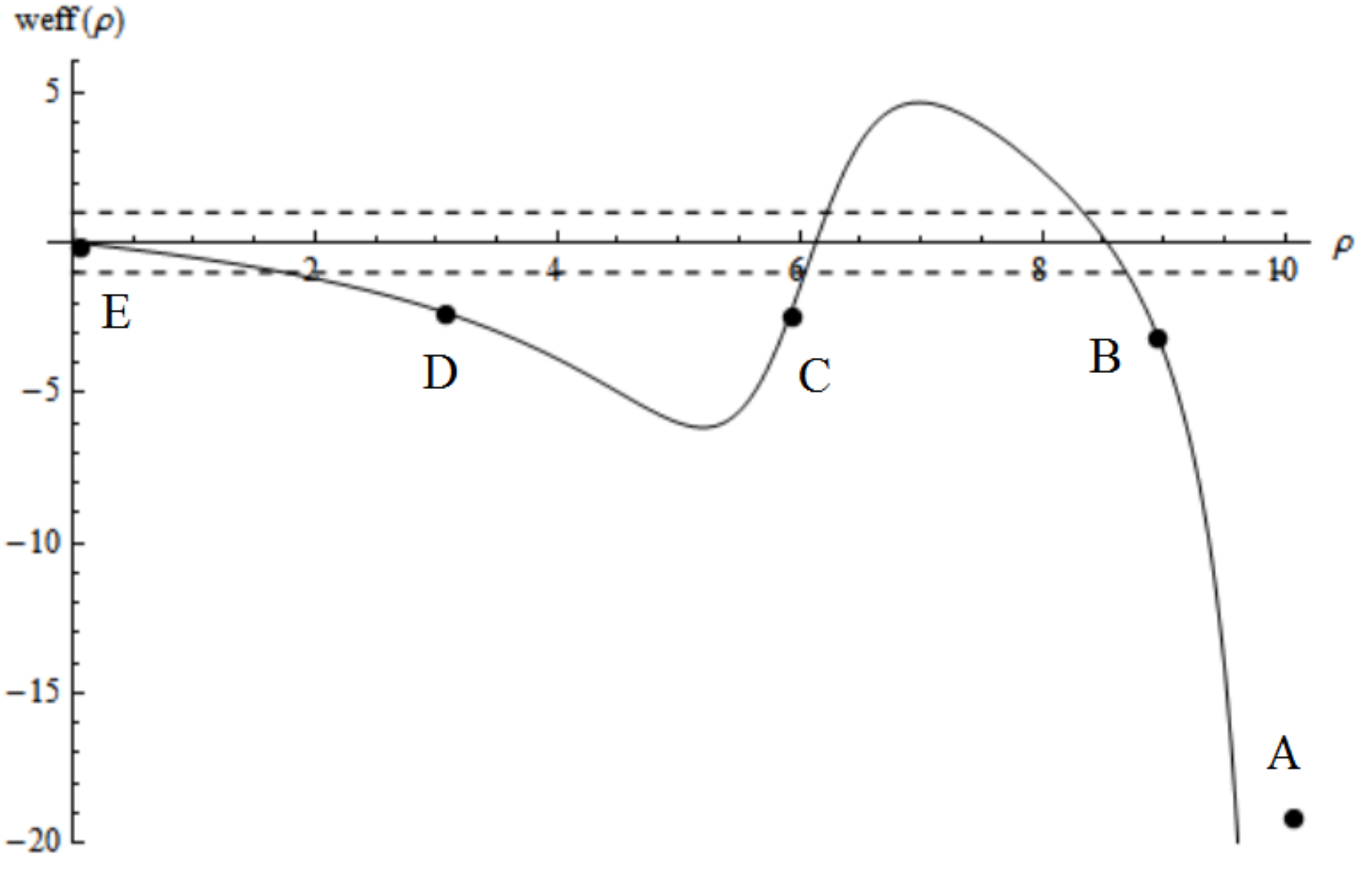

In Figure 2 above, we also created the plot of effective equation of state $w_{eff}(\rho)$ vs. $\rho$ for the M-shaped $f(\rho)$ with the same set of numerical constants as in Figure 2. This effective equation of state $w_{eff}(\rho(t))$ for the cyclic universe model does vary from less than -1 to larger than +1 in a continuous fashion.

---

Our cyclic universe model with W-shaped potential well do incorporate the cosmic early inflation era as well as the cosmic late accelerated expansion, thanks to two downhill slopes (point A to point B and point C to point D in Figure 1). This cyclic universe model is thus different from those in [56,57,58] where there is only one downhill slope because their potential $U(\phi)$ is a U-shaped.

We now define $\tau = t(24\pi)^{1/2}\alpha^{-l/2}\left(\beta^2+\gamma^2\right)^{-m/2}$ and also consider $\tau(\rho)$ as a function of $\rho$ instead of $\rho(\tau)$ as a function of $\tau$. The solution to (3.6) then becomes,

$$\tau(\rho)=\tau(\alpha)+\int_{\rho}^{\alpha}\left(x+p(x)\right)^{-1}\left(x\left(1-\delta^n x^{-n}\right)\left(\left(x^m-\beta^m\right)^2+\gamma^2\right)\left(1-\alpha^{-l}x^l\right)\right)^{-1/2}dx \qquad (3.7)$$

The time, $T$, it takes to go from big-band at point A to big-stop at point E is given by:

$$T=\tau(\delta)-\tau(\alpha)=\int_{\lambda}^{\alpha}\left(x+p(x)\right)^{-1}\left(x\left(1-\delta^n x^{-n}\right)\left(\left(x^m-\beta^m\right)^2+\gamma^2\right)\left(1-\alpha^{-l}x^l\right)\right)^{-1/2}dx \qquad (3.8)$$

We now assume $p = w\rho^r$ ($r \ge 1$). We emphasize that if $\delta = 0$, then the dominant part of integrand (3.8) near $x = 0$ becomes $x^{-3/2}$ so (3.8) diverges. Thus we need a positive but tiny $\delta$ (**no matter how tiny it is**, e.g., $\delta = 10^{-100}m^{-2}$ in metric unit) to obtain a finite $T$ and thus a cyclic universe model. When $n = 1$, we can relate $\delta$ to the cosmology constant $\Lambda$ via $\delta = -(8\pi G)^{-1}c^4\Lambda$ (we recovered the unit of gravitational constant $G$ and speed of light $c$ here). Hence we need a negative but tiny cosmology constant $\Lambda$ to make $T$ finite and the cosmology model cyclic.

In Appendix B, we show that when $l = m = n = r = 1$, (3.7) can be expressed as a linear combination of Elliptical integrals of first kind, third kind, as well as an elementary function $\arctan$.

The optimal values for 7-parameters $(\alpha,\beta,\gamma,\delta,l,m,n)$ in the M-shaped $f(\rho)$ should be determined by the fitting of this model with observational data. This is beyond our current capability.

We now go through our check list to see if we can achieve 5 goals mentioned in the abstract with the cyclic universe model with 7-parameter M-shaped $f(\rho)$ of (3.5).

Goal number (1) producing a non-singular cosmological model $\left(0 < R^2 < \infty\right)$, $\left(0 < R_{\mu\nu}R^{\mu\nu} < \infty\right)$. From (3.6), because $\dot{\rho}^2 \geq 0$ we get $\delta \leq \rho \leq \alpha$. Thus $\rho_{eff}(\rho)(= f(\rho))$ is bounded from above and below in this range. And week energy condition is also satisfied. To show $p_{eff}(\rho)$ is bounded; it is suffice to show that $p(\rho)\partial_\rho f(\rho)$ is bounded. Substituting $p = w\rho^r \ (r \geq 1)$ and (3.5) into $p(\rho)\partial_\rho f(\rho)$, It is straight forward to show that $p(\rho)\partial_\rho f(\rho)$ is bounded from above and below in the range $\delta \leq \rho \leq \alpha$. Thus via (2.5) we achieve goal number (1).

Goal number (2) explaining the cosmic early inflation and late acceleration in a unified fashion. See Figure 1 and the description right after.

Goal number (3) passing the solar system tests. As long as $\alpha, \beta$ large enough and $\delta$ is tiny enough, we have $f(\rho) \approx \rho$ and we can pass the solar system tests.

Goal number (4) unifying the dark matter with dark energy. See Figures 1 and 2 and the description in between.

---

Goal number (5) unifying the regular matter/energy with dark matter/energy in a seamless fashion. Unlike other dark energy (+ regular matter) theory, there is only one material (one energy density $\rho$ and one pressure $p$) in our $f(\rho)$ theory based cyclic universe model. This single material plays both the role of regular matter/energy when needed in the FLRW era and the role of dark matter/energy when needed in other eras in the cyclic universe model. See Figures 2 and 3 and the description in between for details. It is in this sense that we meant we achieve goal number (5). Maybe regular matter/energy (like perfect fluid) and dark matter/energy are just two aspects of the same material. In other words, we have shown the bright side of the dark matter/energy or the dark side of the regular matter/energy (perfect fluid).

The concept of cyclic universe model itself is not new. For example, any cosmology in general and cyclic cosmology in particular, could be reconstructed in $F(R)$ theories of gravity. The corresponding technique is described in [59,60]. Realistic $F(R)$ gravity cosmology model has recently been proposed in [61]. This model can also achieve goals (1) through (4) mentioned above.

One unique thing about the $f(\rho)$ theory based cyclic model of (3.6) is that, the mathematics is very much simplified. Without specifying $p(\rho)$ and solving the Fridmann equations (3.6), we can already illustrate the main features of this cosmological model.

---

# V. The relation between $f(\rho)$ theory and $F(R)$ theories, Chaplygin Gas, NED, etc

## (A) The relation between $f(\rho)$ theory and $F(R)$ theories

In $F(R)$ theory, the Einstein's equations can be cast into a form similar to (2.1) (with $F'(R)\equiv\partial_R F(R)$)[62]:

$$R_{\mu\nu}-\tfrac{1}{2}g_{\mu\nu}R=\frac{8\pi}{F'(R)}\left(T_{\mu\nu}^{(perfect\ fluid)}+T_{\mu\nu}^{(eff)}(R)\right) \tag{5.1}$$

$$T_{\mu\nu}^{(perfect\ fluid)}=(\rho+p)\,u_\mu u_\nu+p\ g_{\mu\nu} \tag{5.2a}$$

$$T_{\mu\nu}^{(eff)}(R)=\frac{1}{16\pi}g_{\mu\nu}\left(F(R)-RF'(R)\right)+\frac{1}{8\pi}\left(\nabla_\mu\nabla_\nu-g_{\mu\nu}g^{\kappa\lambda}\nabla_\kappa\nabla_\lambda\right)F'(R) \tag{5.2b}$$

If we equate (2.1) to (5.1), then we obtain a formal relation between the energy-momentum tensor of $f(\rho)$ theory in (2.2) and that of $F(R)$ theories in (5.2):

$$T_{\mu\nu}\left(\rho,p,f(\rho)\right)=\frac{1}{F'(R)}\left(T_{\mu\nu}^{(perfect\ fluid)}(\rho,p)+T_{\mu\nu}^{(eff)}\left(F(R)\right)\right) \tag{5.3}$$

It remains to be seen if anything significant can be deduced from (5.3).

We would look at the relation between $f(\rho)$ theory and $F(R)$ theories (and Palatini $F(\mathcal{R})$ theory) from a different angle. We can start by looking at their modified actions against the original Hilbert-Einstein action. Let us first compare the corresponding Lagrangians with the corresponding trace equations (with units $8\pi G=c=1$ in this section) for dust ( $p = w\rho$ , $w = 0$ ) in Table 3 below.

Table 3. Comparison of the Lagrangians and trace equations for $f(\rho)$ theory, $F(R)$ theories, and Palatini $F(\mathcal{R})$ theory.

| | spacetime Lagrangian + material Lagrangian | Trace of Einstein's equation $(R=-T)$ |
|---|---|---|
| Einstein | $R-\rho$ (5.4a) | $R=\rho$ (5.4b) |
| $f(\rho)$ theory | $R-f(\rho)$ (5.5a) | $R=4f(\rho)-3\rho\partial_\rho f(\rho)$ (5.5b) |
| $F(R)$ theory[63] | $f(R)-\rho$ (5.6a) | $-3\nabla_\mu\nabla^\mu\left(\partial_R F(R)\right)-R\partial_R F(R)+2F(R)=\rho$ (5.6b) |
| Palatini $F(\mathcal{R})$ theory[64] | $F(\mathcal{R})-\rho$ (5.7a) | $-\mathcal{R}\partial_{\mathcal{R}}F(\mathcal{R})+2F(\mathcal{R})=\rho$ (5.7b) |

We start with (5.4b), the trace of the original Einstein's equation for dust, $R=\rho$. We observe that as long as $R$ and $\rho$ go to infinity at the same speed, the trace equation $R=\rho$ can still be satisfied. We think that this is the cause of the intrinsic singularity.

(i) One way to break up this running away (to infinity) situation is to replace $R=\rho$ with $R=f(\rho)=\rho e^{-b\rho}$ with $\left(b^{-1}\propto\rho_{Planck}\right)$. From this equation, we can immediately obtain $|R|\le b^{-1}$. Consequently $\rho$ is, via $R=f(\rho)=\rho e^{-b\rho}$, also bounded from above and below. If we replace $R=\rho$ with $R=f(\rho)=\rho(1+b\rho)^{-2}$, we can obtain $|R|\le b^{-1}$ as well. Of course, we can not just replace $\rho$ with $f(\rho)$ in the trace of Einstein's equations. We have to do it at the Lagrangian level. Thus we are led to replace Lagrangian of (5.4a) in Einstein's theory, $R-\rho$, with that of (5.5a) in $f(\rho)$ theory, $R-f(\rho)$. The resulting trace equation is (5.5b), $R=4f(\rho)-3\rho\partial_\rho f(\rho)$. Assuming $f(\rho)=\rho e^{-b\rho}$ or $f(\rho)=\rho(1+b\rho)^{-2}$, it is straight forward to show from (5.5b) that $R$ is bounded from above and below and so is $\rho$.

(ii) Another way to break up this running away (to infinity) situation is to replace Lagrangian of (5.4a) in Einstein's theory, $R-\rho$, with that of (5.7a) in Palatini $F(\mathcal{R})$ theory, $F(\mathcal{R})-\rho$. The resulting trace equation is (5.7b), $-\mathcal{R}\partial_{\mathcal{R}}F(\mathcal{R})+2F(\mathcal{R})=\rho$. Assuming $F(\mathcal{R})=R e^{-\frac{1}{2}\beta^2\mathcal{R}^2}$ with $\left(\beta\propto l_{Planck}\right)$, it is straight forward to show from (5.7b) that $0\le\rho\le\beta^{-1}$, i.e., bounded from above and below. Consequently $\mathcal{R}$ is, via $-\mathcal{R}\partial_{\mathcal{R}}F(\mathcal{R})+2F(\mathcal{R})=\rho$, also bounded from above and below. If we select $F(\mathcal{R})=\mathcal{R}\left(1+\beta^2\mathcal{R}^2\right)^{-1}$, we can show that $\rho$ and $\mathcal{R}$ are bounded from above and below as well.

Because of the higher order derivative term in (5.6b), a similar analysis is doable but less straight forward.

### (B) The relation between $f(\rho)$ theory and Nonlinear ElectroDynamics (NED)

We are intrigued by the nonsingular exact black hole solutions with Nonlinear ElectroDynamics (NED) [65,66,67]. As a matter of factor, the inspiration to both the original authors of **[39, 65]** can be traced back to the famous Born-Infeld theory [68].

The Lagrangian $f(\rho)$ and the NED Lagrangian $L\left(F_{\mu\nu}F^{\mu\nu}\right)$ can be used material Lagrangian as the source in Hilbert-Einstein like action $S=(16\pi)^{-1}\int\left(R-f(\rho)\right)\sqrt{-g}\,d^4x$ and $S=(16\pi)^{-1}\int\left(R-L\left(F_{\mu\nu}F^{\mu\nu}\right)\right)\sqrt{-g}\,d^4x$. One major difference between these two theories is that the energy-momentum tensor generated from $L\left(F_{\mu\nu}F^{\mu\nu}\right)$ is traceless while as the energy-momentum tensor generated from $f(\rho)$ (see (2.1)) is not.

Recall that in Born-Infeld theory the action is given by $S=\int\left(L\left(F_{\mu\nu}\right)-j_\mu A^\mu\right)\sqrt{-\eta}\,d^4x$ with $L\left(F_{\mu\nu}\right)=b^2\left(1-\left(\det\left(\eta_{\mu\nu}+b^{-1}F_{\mu\nu}\right)\right)^{1/2}\right)$. The square root function is used to break up the running away to infinity situation (at the center of a charged particle). Thus we would guess that $L\left(F_{\mu\nu}\right)=\frac{1}{4}\left(F_{\mu\nu}F^{\mu\nu}\right)\left(1+b^{-4}\left(F_{\mu\nu}F^{\mu\nu}\right)^2\right)$ might do a similar job in breaking up the running away situation.

### (C) The relation between $f(\rho)$ theory and Chaplygin Gas

The Chaplygin gas is an exotic perfect fluid (a kind of dark energy) with equation of state [69,70,71]:

$$p(\rho)=-A\rho^{-\alpha} \qquad (0<A, 0<\alpha\leq 1) \tag{5.9}$$

It is used to explain the cosmic late accelerated expansion.

Notice that the equation of state behaves like:

$$w(\rho)=p(\rho)/\rho=-A\rho^{-(1+\alpha)}<0 \tag{5.10a}$$
$$\partial_\rho w(\rho)=p(\rho)/\rho=(1+\alpha)A\rho^{-(2+\alpha)}>0 \tag{5.10b}$$

From Figure 3, we deduce that $w_{eff}(\rho)$ near $(\rho=\beta)$ (point C) behaves like a Chaplygin gas.

The problem with Chaplygin gas of (5.9) is that it diverges at $\rho=0$ in the range $\rho\in[0,\infty)$.

The $f(\rho)$ based cyclic universe model of (3.5) does not have this kind of divergence problem.

# VI. Conclusion

We considered FLRW cosmological model for perfect fluid (with $\rho\geq 0$ as the energy density) in the frame work of the $f(\rho)$ modified theory of gravity. With an M-shaped function $f(\rho)$, we achieved, like in $F(R)$ modified theory of gravity, the following 4 specific goals, (1) producing a non-singular cosmological model $\left(0<R^2<\infty\right)$, $\left(0<R_{\mu\nu}R^{\mu\nu}<\infty\right)$; (2) explaining the cosmic early inflation and late acceleration in a unified fashion; (3) passing the solar system tests; (4) unifying the dark matter with dark energy.

In addition, we also achieve goal number (5): unifying the regular matter/energy with dark matter/energy in a seamless fashion in $f(\rho)$ theory and goal number (6): simplifying the Einstein's equations (in comparison to $F(R)$ theory).

We would like to emphasize that in our $f(\rho)$ theory based cyclic universe model, the single material (energy density $\rho$) plays both the role of regular matter/energy when needed in the FLRW era and the role of dark matter/energy when needed in other eras in the cyclic universe. Thus we guess that the regular matter/energy (like perfect fluid) and dark matter/energy might just be two aspects of the same material. In other words, we have probably shown the bright side of the dark matter/energy or the dark side of the regular matter/energy (perfect fluid).

The apparent unification of the regular matter/energy with dark matter/energy in a seamless fashion and simplification of Einstein's equations (Friedman's equations) in $f(\rho)$ theory are probably the two interesting benefits of using $f(\rho)$ theory.

# Acknowledgement

During the development of this work, we have greatly benefited from the stimulating discussions with Dr. Jie Qing, Dr. J. E. Hearst, Dr. W. M. McClain, Dr. R. A. Harris, Dr. R. P. Lin, Dr. Lei Xu, Dr. Zixiang Zhou, Dr. Ying-Qiu Gu, Dr. Ru-keng Su, Dr. Bin Wang, Dr. Jinglan Sun, and Dr. W. Liu.

We would also appreciate the feedback from Dr. S. D. Odintsov, Dr. S. Nojiri, Dr. E. Elizalde, Dr. T. P. Sotiriou, Dr. V. Faraoni, and Dr. K. Lake.

# Appendix A. Derivation of the Energy-Momentum Tensor from Lagrangian $f(\rho)$.

We now follow the standard text book procedure (cf. Hawking and Ellis [72]) to derive the energy-momentum density tensor using the material Lagrangian $f(\rho)$ for the ideal fluid. We denote $\sigma$ the mass density, $p(\sigma)$ the pressure, $\varepsilon(\sigma)$ the internal energy density, and $\rho(\sigma)$ the (total) energy density of the ideal fluid. We remark that in [72] symbol $\rho$ represents mass density (so it is equivalent to our $\sigma$ here) and symbol $\mu$ represents (total) energy density (so it is equivalent to our $\rho$ here).

The energy-momentum tensor is defined as:

$$\frac{1}{2}\sqrt{-g}T^{ab} = \frac{\partial\left(\sqrt{-g}L_{material}\right)}{\partial g_{ab}} \tag{A.1}$$

In [72], the material Lagrangian is chosen as:

$$L_{material}(\sigma) = \rho \equiv \sigma(1+\varepsilon). \tag{A.2}$$

Here we choose:

$$L_{material}(\sigma) = f(\rho). \qquad \rho \equiv \sigma(1+\varepsilon) \tag{A.3}$$

Substitution of (A.3) into (A.1), we have:

$$\frac{1}{2}\sqrt{-g}T^{\mu\nu} = f(\rho)\frac{\partial\left(\sqrt{-g}\right)}{\partial g_{\mu\nu}} + \sqrt{-g}\partial_\rho f(\rho)\frac{\partial\left(\sigma(1+\varepsilon)\right)}{\partial\sigma}\frac{\partial\sigma}{\partial g_{\mu\nu}} \tag{A.4}$$

The derivation may be simplified by noting that the conservation of the current $j^\nu = \sigma u^\nu$ may be expressed as:

$$\nabla_\nu j^\nu = \frac{1}{\sqrt{-g}}\partial_\nu\left(\sqrt{-g}\, j^\nu\right) = 0 \tag{A.5}$$

Given the flow lines, the conservation equations determine $j^\nu$ uniquely at each point on a flow line in terms of its initial value at some given points on the same flow line. Therefore $\sqrt{-g}\, j^\nu$ is unchanged when the metric is varied. Since

$$\sigma^2 = g^{-1}\left(\sqrt{-g}\, j^\mu\right)\left(\sqrt{-g}\, j^\nu\right) g_{\mu\nu} \tag{A.6}$$

So

$$2\sigma\delta\sigma = \left( j^\mu j^\nu - g_{\eta\lambda} j^\eta j^\lambda g^{\mu\nu}\right)\delta g_{\mu\nu} \tag{A.7}$$

Substitution of $j^\nu = \sigma u^\nu$ into (A.7) and realizing that $g_{\eta\lambda}u^\eta u^\lambda \equiv -1$, we obtain:

$$\delta\sigma = \tfrac{1}{2}\sigma\left(u^\mu u^\nu + g^{\mu\nu}\right)\delta g_{\mu\nu} \tag{A.8a}$$

$$\frac{\partial\sigma}{\partial g_{\mu\nu}} = \tfrac{1}{2}\sigma\left(u^\mu u^\nu + g^{\mu\nu}\right) \tag{A.8b}$$

We also have:

$$\frac{\partial\left(\sqrt{-g}\right)}{\partial g_{ab}} = -\tfrac{1}{2}\sqrt{-g}\, g^{ab} \tag{A.9}$$

We recall that the combined First law and Second law of thermodynamics in the rest frame is given by:

$$Tds = d\varepsilon + P\, d\left(1/\sigma\right) \tag{A.10}$$

In (A.10), $T$ is temperature and $s$ is the entropy. For the isentropic perfect fluid ($ds = 0$) we thus obtain:

$$\frac{\partial\varepsilon}{\partial\sigma} = \frac{P}{\sigma^2} \tag{A.11}$$

Substitution of (A.8b), (A.9), and (A.11) into (A.4), we finally obtain:

$$T^{\mu\nu} = \left[(\rho + p)\partial_\rho f(\rho)\right] u^\mu u^\nu + \left[(\rho + p)\partial_\rho f(\rho) - f(\rho)\right] g^{\mu\nu} \tag{A.12a}$$

and

$$T_{\mu\nu} = \left[(\rho + p)\partial_\rho f(\rho)\right] u_\mu u_\nu + \left[(\rho + p)\partial_\rho f(\rho) - f(\rho)\right] g_{\mu\nu} \tag{A.12b}$$

We shall call a perfect fluid isentropic if the pressure $p$ is independent of entropy $s$ and is a function of (total) energy density only, $p = p(\rho)$. In this case one can introduce a conserved mass density $\sigma$ and an internal energy $\varepsilon$ and derive the energy-momentum tensor from a Lagragian $\rho \equiv \sigma(1+\varepsilon)$ as in [72] or a Lagragian $f(\rho) \equiv f(\sigma(1+\varepsilon))$ as here.

Eq.(A.12b) is equivalent to (2.2)-(2.3) of the main text. So we accomplish the first task in this appendix.

The second task we want to accomplish is to convert the energy-momentum tensor in [40] into the same form as in (A.12). We remark that in [40] symbol $\rho$ represents mass density (so it is equivalent to our $\sigma$ here).

In our notation the energy-momentum tensor (in the absence of torsion) is given by (cf. (3.4h) of [40]):

$$T^{\mu\nu} = \sigma\partial_\sigma F(\sigma)\, u^\mu u^\nu + \left(\sigma\partial_\sigma F(\sigma) - F(\sigma)\right) g^{\mu\nu} \tag{A.13}$$

Notice that (A.13) looks different from (A.12b) in the following ways: (i) the differentiation is wrt to mass density $\sigma$ [in (A.12b), the differentiation is wrt to (total) energy density $\rho$]; (ii) the pressure $p$ does not appear explicitly [in (A.12b), the pressure $p$ does appear explicitly]. We remark that (A.13) is not suitable for practical calculation because pressure $p$ does not appear explicitly.

In [40], $F(\sigma)$ is eventually chosen as $F(\sigma) \equiv \sigma(1+\varepsilon)$. In this appendix, we identifying $F(\sigma) = f(\rho)$ [while remembering $\rho \equiv \sigma(1+\varepsilon)$]. Using chain rule $\partial_\sigma = (\partial\rho/\partial\sigma)\partial_\rho$, and the pressure definition (A.11), we obtain:

$$\begin{aligned}
\sigma\partial_\sigma F(\sigma) &= \sigma\partial_\sigma f(\rho) = \sigma\left(\partial_\rho f(\rho)\right)\frac{\partial\rho}{\partial\sigma} \\
&= \sigma\left(\partial_\rho f(\rho)\right)\frac{\partial(\sigma + \sigma\varepsilon)}{\partial\sigma} = \sigma\left(\partial_\rho f(\rho)\right)\left(1 + \varepsilon + \sigma\frac{\partial\varepsilon}{\partial\sigma}\right) \\
&= \left(\partial_\rho f(\rho)\right)\left(\sigma(1+\varepsilon) + \sigma^2\frac{\partial\varepsilon}{\partial\sigma}\right) = \left(\partial_\rho f(\rho)\right)(\rho + p)
\end{aligned} \tag{A.14}$$

Substitution of (A.14) into (A.13), we find out that the result is identical to (A.12a).

# Appendix B. Conversion of an elliptical integral into Legendre standard form

The elliptical integral we want to convert, (3.7) with $\left(l=m=n=r=1\right)$ and $\left(x=\rho,\sigma=\tau\sqrt{1+w}\right)$, can be rewritten as:

$$d\sigma=-\frac{dx}{x\sqrt{P(x)}} \tag{B.1a}$$

$$P(x)=(\alpha-x)(x-\delta)(x-\beta-i\gamma)(x-\beta+i\gamma) \tag{B.1b}$$

Following the standard procedures [73], we can convert (B.1) to:

$$-\frac{dx}{x\sqrt{P(x)}}=-\frac{1}{(A-B)}\frac{(Bs-A)ds}{(s-1)\sqrt{Q(s)}} \tag{B.2a}$$

$$Q(s)=\left(b_1s^2+c_1\right)\left(b_2s^2+c_2\right) \tag{B.2b}$$

Where

$$s=\frac{x-A}{x-B} \tag{B.3a}$$

$$A=\frac{\alpha\delta-\left(\beta^2+\gamma^2\right)+\sqrt{\left((\alpha-\beta)^2+\gamma^2\right)\left((\delta-\beta)^2+\gamma^2\right)}}{\alpha+\delta-2\beta} \tag{B.3b}$$

$$B=\frac{\alpha\delta-\left(\beta^2+\gamma^2\right)-\sqrt{\left((\alpha-\beta)^2+\gamma^2\right)\left((\delta-\beta)^2+\gamma^2\right)}}{\alpha+\delta-2\beta} \tag{B.3c}$$

$$b_1=\frac{-2B+\alpha+\delta}{2A-2B} \qquad c_1=-\frac{-2A+\alpha+\delta}{2A-2B} \tag{B.3d}$$

$$b_2=\frac{B-\beta}{A-B} \qquad c_2=\frac{-A+\beta}{A-B} \tag{B.3e}$$

We next multiply the numerator and denominator of RHS of (B.2a) by $(s+1)$ and obtain:

$$-\frac{(Bs-A)ds}{(A-B)(s-1)\sqrt{Q(s)}}=-\frac{Bds}{(A-B)\sqrt{Q(s)}}-\frac{ds}{\left(1-s^2\right)\sqrt{Q(s)}}-\frac{d\left(s^2\right)}{2\left(1-s^2\right)\sqrt{Q(s)}} \quad \text{(B.4)}$$

For our case we have $b_1>0, c_1<0, b_2<0, c_2<0$. So we can rewrite $Q(s)$ as $Q(s)=(b_1)(-b_2)\left(a^2-s^2\right)\left(s^2+b^2\right)$, where $b^2=(-c_2)/(-b_2)$, $a^2=(-c_1)/b_1$. This belongs to case (4) [68]: $s<a$. We now define $s^2=a^2\left(1-t^2\right)$, $k^2=a^2/\left(a^2+b^2\right)$. The first term on the RHS of (B.4) becomes the elliptical integral of the first kind:

$$-\frac{Bds}{(A-B)\sqrt{Q(s)}}=\left[\frac{B}{(A-B)\sqrt{|b_2c_1|+|b_1c_2|}}\right]\frac{dt}{\sqrt{\left(1-t^2\right)\left(1-k^2t^2\right)}} \quad \text{(B.5)}$$

The second term on the RHS of (B.4) becomes the elliptical integral of the third kind:

$$-\frac{ds}{\left(1-s^2\right)\sqrt{Q(s)}}=\left[\frac{1}{\left(1-a^2\right)\sqrt{|b_1c_2|+|b_2c_1|}}\right]\frac{dt}{\left(1-nt^2\right)\sqrt{\left(1-t^2\right)\left(1-k^2t^2\right)}} \quad \text{(B.6)}$$

$$\left(n=a^2/\left(a^2-1\right)\right)$$

The third term on the RHS of (B.4) becomes the one of the elementary functions, arctan:

$$-\frac{d\left(s^2\right)}{2\left(1-s^2\right)\sqrt{Q(s)}}=\left[-\frac{1}{\sqrt{|b_1b_2|}\sqrt{\left(1-a^2\right)\left(1+b^2\right)}}\right]d\left(\arctan\left(\frac{\sqrt{\left(b^2+1\right)\left(a^2-s^2\right)}}{\sqrt{\left(1-a^2\right)\left(b^2+s^2\right)}}\right)\right) \quad \text{(B.7)}$$

If $\alpha+\delta-2\beta=0$, (B.3) will not apply. In this case we have $A=\frac{1}{2}(\alpha+\delta)=\beta$, $B=\infty$. We can define $s=x-A$ and obtain $ds=dx$ and

$$P(x)=P(s+A)=\left(\beta^2+\alpha\delta-s^2\right)\left(s^2+\gamma^2\right)=Q(s), \quad \text{(B.8)}$$

Thus we converted (B.1) into (B.2). The rest of treatment is similar to (B.4) through

(B.7).